\documentclass[11pt,twoside]{article}
\usepackage{asp2010}

\resetcounters

\begin{document}

\title{The GREGOR solar telescope on Tenerife}
\author{W. Schmidt$^1$, O. von der L\"uhe$^1$, R. Volkmer$^1$, C. Denker$^2$, S.K. Solanki$^3$, H. Balthasar$^2$,  N. Bello Gonzalez$^1$,T. Berkefeld$^1$, M. Collados$^4$ A. Hofmann$^2$, F. Kneer$^5$, A. Lagg$^3$, K. Puschmann$^2$, D. Schmidt$^1$, M. Sobotka$^6$, D. Soltau$^1$, and K.G. Strassmeier$^2$
\affil{$^1$Kiepenheuer-Institut f\"ur Sonnenphysik, Sch\"oneckstra\ss e 6, 79104 Freiburg, Germany}
\affil{$^2$Leibniz-Institut f\"ur Astrophysik, An der Sternwarte 16, 14482 Potsdam, Germany}
\affil{$^3$Max-Planck-Institut f\"ur Sonnensystemforschung, Max-Planck-Stra\ss e 2, 37191 Katlenburg-Lindau, Germany}
\affil{$^4$Instituto de Astrof\'\i sica de Canarias, V\'\i a Lactea, 38200 La Laguna, Spain}
\affil{$^5$Institut f\"ur Astrophysik der Georg-August Universit\"at G\"ottingen, Friedrich-Hund-Platz 1, 37077 G\"ottingen, Germany}
\affil{$^6$Astronomical Institute, Academy of Sciences, Fri\v{c}ova 298, 25165 Ond\v{r}ejov, Czech Republic}}

\begin{abstract}
2011 was a successful year for the GREGOR project. The telescope was finally completed in May with the installation of the 1.5-meter primary mirror. The installation of the first-light focal plane instruments was completed by the end of the year. At the same time, the preparations for the installation of the high-order adaptive optics were finished, its integration to the telescope is scheduled for early 2012. This paper describes the telescope and its instrumentation in their present first-light configuration, and provides a brief overview of the science goals of GREGOR. 
\end{abstract}

\section{Introduction}

\begin{figure}[!ht]
\includegraphics[width=\textwidth]{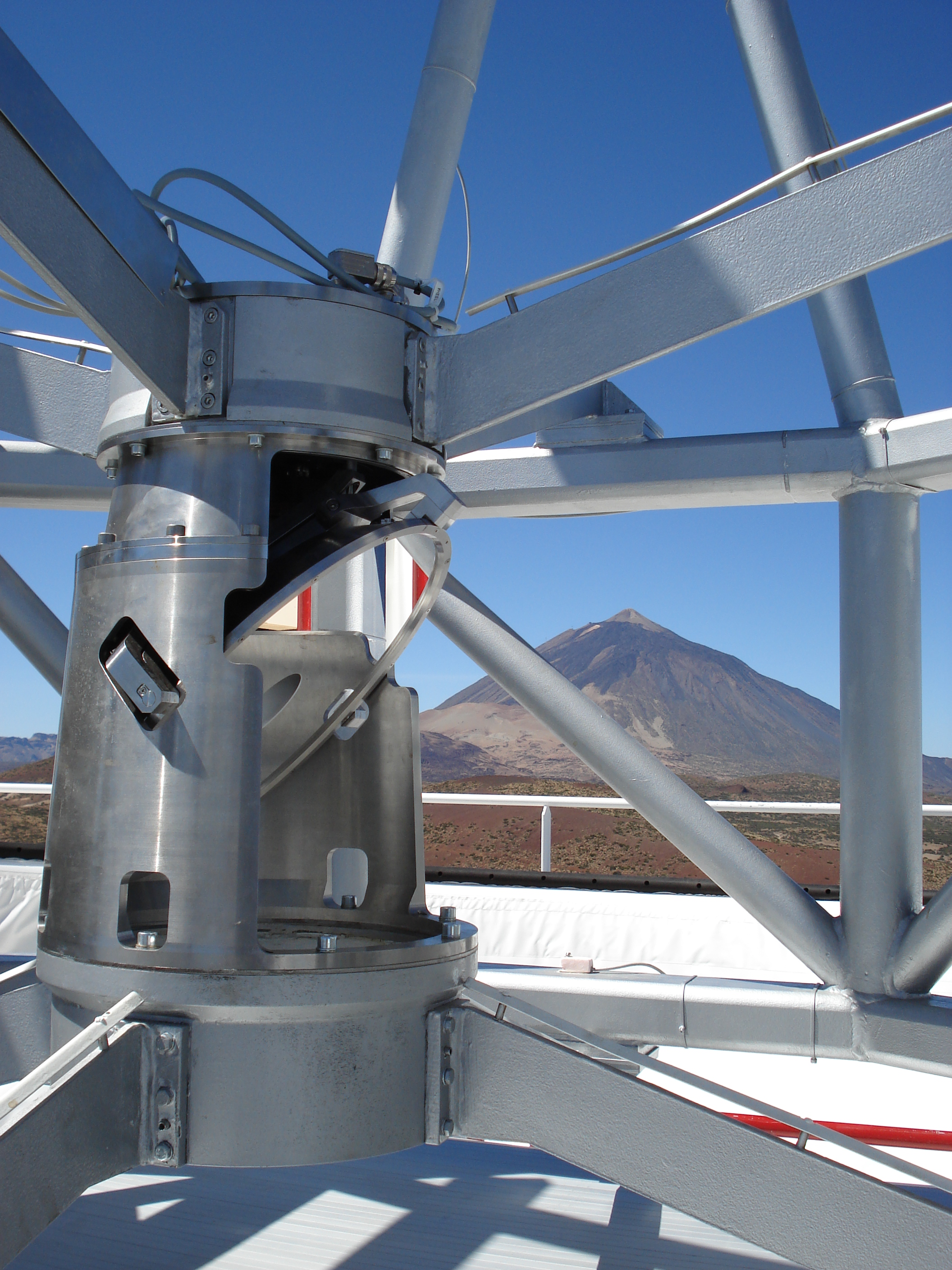}
\caption{\label{fig_tel}Partial view of the GREGOR telescope structure with the folding mirror \textsf{M4}. The light beam coming from \textsf{M2} passes through the central hole of \textsf{M4} down to \textsf{M3}, and the return beam is then reflected by \textsf{M4} into the elevation axis. The secondary focal plane \textsf{F2} is located just above \textsf{M4}, where also the polarization calibration unit is placed.}
\end{figure}

Important changes in the design philosophy of solar telescopes occurred during the last decade. In the past, all major solar telescopes with apertures above 50 cm had been designed as evacuated systems in order to avoid mirror seeing, i.e., local turbulence caused by heated air above the telescope primary or elsewhere in the optical path. The concept of vacuum telescopes is not applicable to telescopes with apertures significantly above one meter. Therefore, the two newest solar telescopes with apertures above one meter, the German GREGOR 1.5-meter telescope at the Observatorio del Teide, and the 1.6-meter NST \citep{Cao+etal2010} at Big Bear Observatory in the US, are designed and built as open telescopes. Their optical designs are pathfinders for the development of the next-generation telescopes of the 4-meter class. 

Open solar telescopes require significant effort to avoid mirror seeing caused by the absorbed thermal energy in the primary mirror. The NST and GREGOR use different approaches. While GREGOR has a retractable dome and uses wind flushing of the telescope area to avoid or remove near-telescope turbulence, the NST has a closed dome and uses active venting of the dome area to prevent dome seeing. The NST optical off-axis design is quite similar to that of the Advanced Technology Solar Telescope (ATST), whereas GREGOR with its on-axis optics can be considered a pathfinder design for the European Solar Telescope (EST). 
The GREGOR project started in 2000, and it experienced significant delays, mainly caused by the fabrication of the primary mirror that was planned to be made from silicon carbide (CESIC) \citep{Kroedel2006}. This innovative approach would have simplified both the mirror mounting, thanks to the high stiffness of CESIC, and it would have simplified the necessary cooling of the mirror surface. Unfortunately, the manufacturing of the CESIC mirror was not successful and finally GREGOR was equipped with a light-weighted Zerodur mirror. At the end of 2011, the telescope is now ready for scientific use, with three first-light instruments installed. These instruments are briefly described in Sect. \ref{instruments}

\section{Science objectives of GREGOR}
Science objectives laid down in the original proposal:
\begin{itemize}
\item \textbf{The interaction between convection and the magnetic field in the photosphere.} This topic includes the structure and dynamics of sunspots with their complex flow, and magnetic field patterns, especially in the penumbra and umbra. In the latter umbral dots are embedded in the cool and strongly magnetized plasma. In the quiet photosphere, the most important themes are the emergence and removal of magnetic flux and the interaction of the small-scale magnetic flux concentrations with the convective motion -- the granulation -- at the surface.

\item \textbf{The solar magnetism and its role for solar variability.} During an 11-year activity cycle, the solar constant varies by about 0.1 \%. This small number holds for the integrated light. In the UV, the variation is much stronger. The increase of the total brightness at activity maximum is caused by small-scale faculae with angular sizes in the order of 0.1\arcsec. Measurements of the strength and the dynamics of the small-scale magnetic field in the \emph{quiet Sun} and its relation to the faculae will help to understand the mechanisms that are responsible for the solar variability. In addition, these measurements will clarify if local dynamo mechanisms are acting in the solar photosphere.

\item \textbf{The enigmatic heating mechanisms of the chromosphere.} There has been substantial progress in the observation of chromospheric structure and dynamics, and some measurements of the chromospheric magnetic field have been made. Nevertheless, the relative importance of the magnetic field and acoustic waves for the heating of the chromosphere is still under intense discussion. The time scales of chromospheric dynamics are short, the magnetic field is small, and the light level is low: these are the ingredients of an observational challenge. Sequences with a cadence of a few seconds and high spatial and spectral resolution are needed to identify the governing heating processes.
\item \textbf{Search for solar twins.} Stellar statistics indicate the existence of about one billion G2 stars in our galaxy. To date, only two solar twins have been identified. We therefore propose a large spectroscopic survey to search for solar twins.
\end{itemize}

The last decade has brought considerable progress for these topics, thanks to high-resolution data from the Swedish Solar Telescope \citep{Scharmer+etal2003}, from the Solar Optical Telescope onboard the Japanese HINODE satellite \citep{Kosugi+etal2007}, and from the stratospheric \textit{Sunrise }telescope \citep{Schmidt+etal2010, Solanki+etal2010, Barthol+etal2011}. Many of the new findings have helped to refine the science questions, and therefore the themes listed above are still valid and up-to-date. As an example, the detection of horizontal magnetic fields almost everywhere in the quiet Sun by \citet{Lites+etal2007} has stimulated the research about the magnetism of the Sun outside active regions. \citet{Borrero+etal2011} show that the accurate measurement of the orientation of the magnetic field in the quiet Sun needs a signal-to-noise ratio much higher than that of the HINODE-SOT measurements. At the spatial resolution of the HINODE-SOT, GREGOR will provide the tenfold photon flux, hence an increases of the signal-to-noise ratio by a factor of three. This, together with the possibility to measure in different layers of the solar atmosphere, will allow a better understanding of the properties of the magnetic field in the solar atmosphere.

\begin{figure}[!ht]

\includegraphics[width=\textwidth]{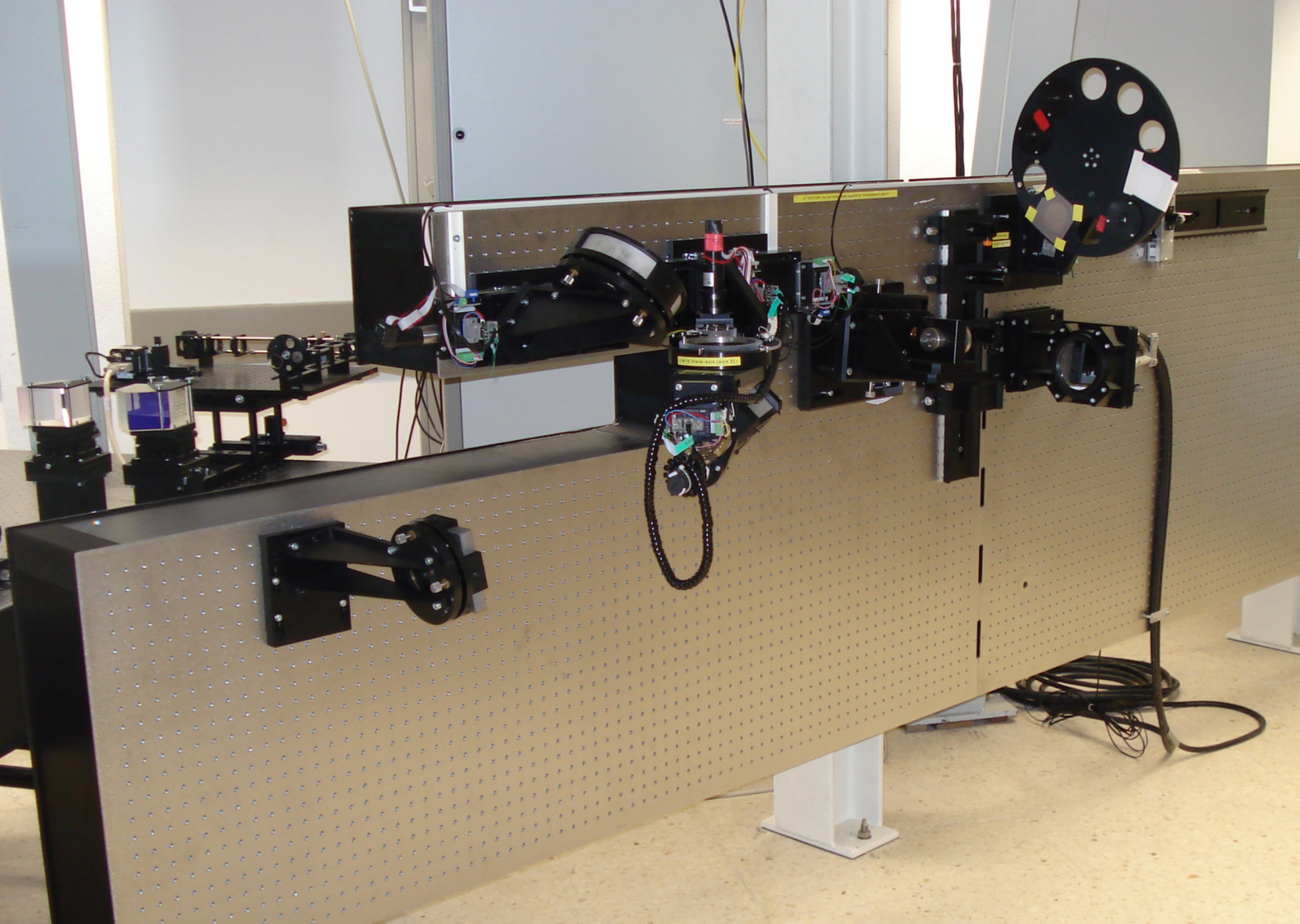}
\caption{\label{fig_ao}Optical bench of the GREGOR adaptive optics system. Part of the unused space is foreseen for the MCAO system. The deformable mirror is located below the filter wheel on the right.}
\end{figure}

\subsection{The importance of large telescope apertures}
 The outer layers of the Sun show intensity, velocity, and magnetic field variations that cover more than five orders of magnitude in size, from one solar radius (meridional flow, differential rotation), down to a few tens of kilometers (magnetic flux concentrations). With existing telescopes of the one-meter class, we are now able to observe structures down to 100 km, and, with restrictions, begin to measure flows and magnetic fields on that same scale.  For precise and accurate measurements of the structure and dynamics of the chromosphere, an aperture of one meter is by far not sufficient, since the outer atmosphere is probed with very strong spectral lines with a tenth or less of the photospheric intensity. For a sufficiently large signal-to-noise ratio (or photometric accuracy), an increase of the photon flux is needed.  GREGOR will provide a substantial increase in photon flux and hence enable high-quality measurements of the magnetic field in the chromosphere. The increase of photon flux with increasing aperture holds for a fixed area on the Sun (or solid angle seen from the telescope). This means that observing an area of 100$\times$100 km$^2$ on the Sun with a 1.5-meter telescope delivers four times  more photons to the focal plane compared to a 70-cm telescope (such as the VTT).

\begin{figure}[!ht]
\includegraphics[width=\textwidth]{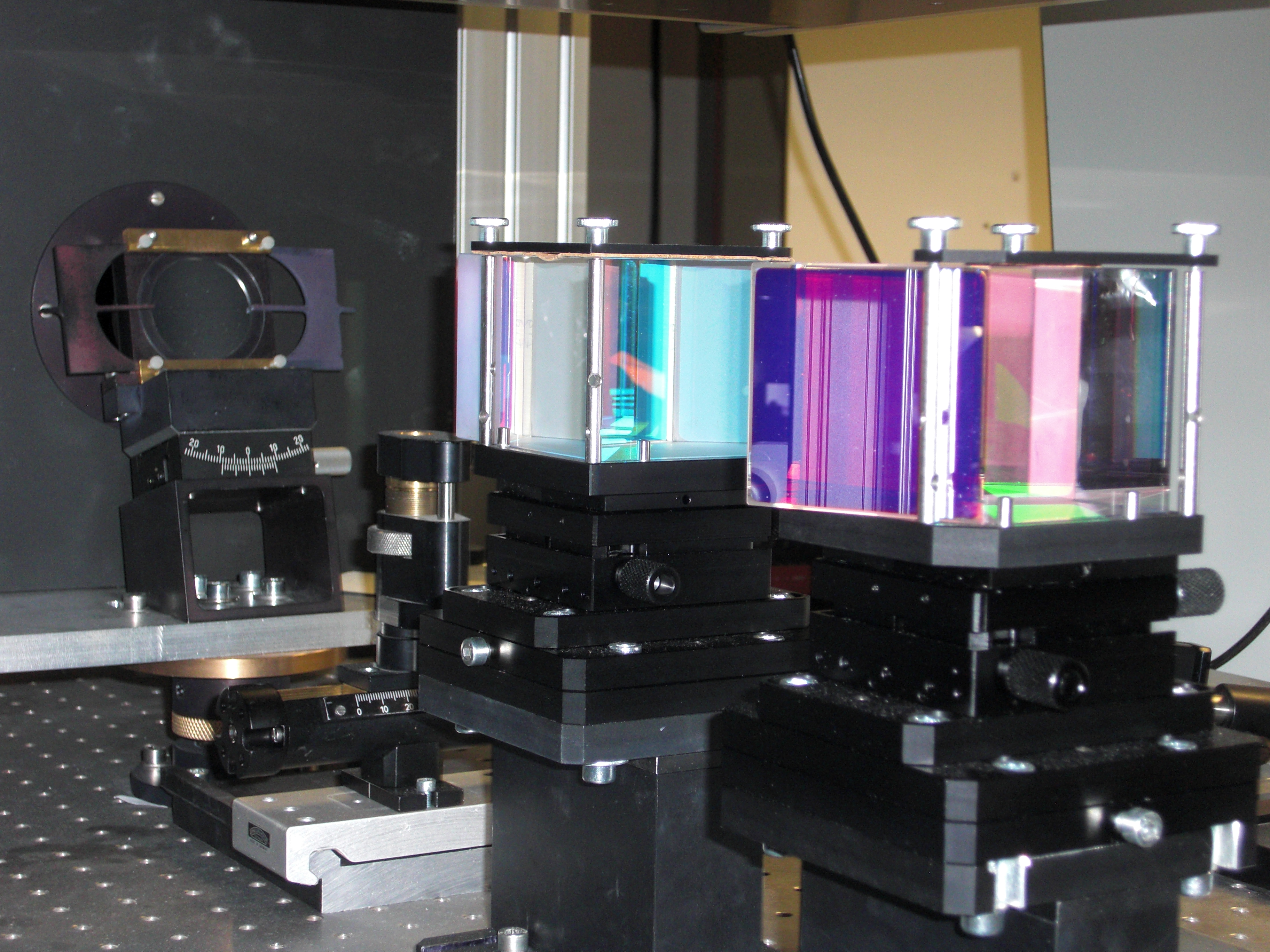}
\caption{\label{fig_bs}The first of these beam splitters sends a fraction of the incoming light to the wavefront sensor of the adaptive optics. The second beam splitter transmits the infrared light ($\lambda \geqslant $ 850 nm) to the infrared spectrograph (Sect. \ref{GRIS}) and reflects all shorter wavelengths to the GFPI (Sect. \ref{GFPI}). The entrance slit of the spectrograph is seen in the background.}
\end{figure}

\section{Telescope and enclosure}
GREGOR was designed as an open telescope, and very early in the project, a decision was made to combine this design with a retractable dome, in order to maximize air flushing of the telescope area by wind. The design of the dome followed the concept used for the 40-cm Dutch Open Telescope \citep{Rutten+etal1999}, but for a much larger telescope. GREGOR is mounted on the top floor of the building at a height of 22 meters above ground. The retractable dome has a diameter of 9 meters. It is based on a steel bar construction, covered with two layers of polyester fabric. The inner layer serves as thermal insulation to prevent water condensation on the steel parts. A Teflon coating on the outside helps to avoid the build-up of an ice layer during periods of snow storms. During observations, the dome will be completely open to allow the ambient air to move across the platform and through the telescope structure, in order to provide natural cooling of the telescope and its immediate environment \citep[see][]{Sliepen2010}.
\subsection{Optical design}
\label{design}
GREGOR is designed as a Gregory system with three imaging mirrors, building upon heritage of the LEST telescope design \citep{Engvold+Andersen1990}. The primary mirror has a diameter of 150 cm and a focal length of 2.5 m. At the location of the primary image, a cooled field stop deflects most of the sunlight out of the telescope, and transmits a circular field-of-view with a diameter of 200\arcsec\ through a central hole. A secondary image is formed near the intersection of the optical and the elevation axis of the telescope. At this place, the (removable) polarization calibration optics is located as well as a folding flat, M4 (see Fig.~\ref{fig_tel}), which directs the light beam into the elevation axis. and further to the observing room one floor below the telescope platform \citep{vonderLuehe+etal2000}. An adaptive optics system is an integral part of the optical design and is located between the Coud\'e focus and the focal plane instruments on an optical bench in the observing room (Fig. \ref{fig_ao}). \citet{Volkmer2006} provide details about the telescope and its optical design.

\begin{figure}[!ht]
\includegraphics[width=\textwidth]{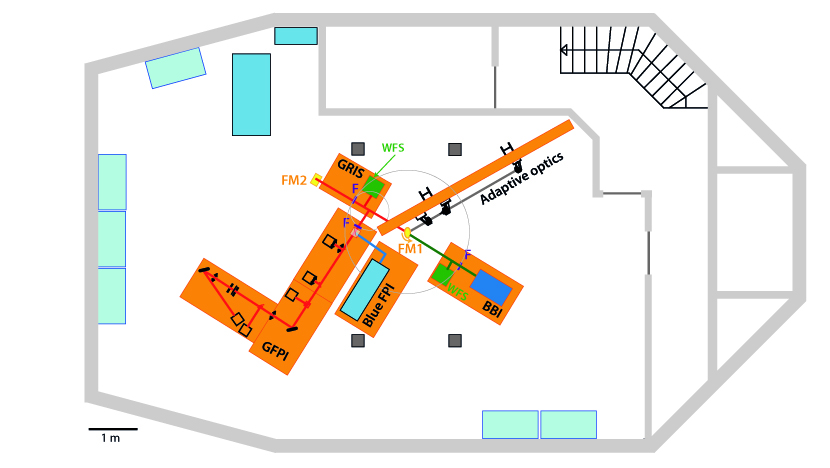}
\caption{\label{gregorlab}Top view of the observing room. The beam coming from the adaptive optics bench can be redirected to different optical tables via the rotating fold mirror FM1. FM2 reflects the light to the infrared grating spectrograph (GRIS) that is located one floor below. GRIS and the visible spectro-polarimeter (GFPI) are fed simultaneously via a dicroic beam splitter. A blue channel of the GFPI is foreseen as a future extension of the GFPI. The broad-band imager (BBI) is presently used as a stand-alone instrument. }
\end{figure}

\subsection{Adaptive optics}  
A low-order adaptive optics system that had been foreseen for the first period of observations with GREGOR has very recently been replaced with a high-order system with 196 actuators. The closed-loop bandwidth of the system (0dB) is 130 Hz. The AO uses a classical Shack-Hartmann wave-front sensor. The number of sub-apertures is 156, each with an aperture of 10 cm.   In the very near future, a multi-conjugate adaptive optics (MCAO) system will be installed and tested at the telescope \citep{Berkefeld+etal2010}. Since the GREGOR adaptive optics operates at visible wavelengths, the isoplanatic patch is quite small, so that an MCAO is desirable. The development of such an MCAO system started at KIS already some years ago \citep{vonderLuehe+etal2005, Berkefeld2006}. Presently, the setup is operational at the KIS, and it is foreseen to install and test the system at the GREGOR telescope in 2012. 

\section{Focal-plane instruments}
\label{instruments}

\begin{figure}[!ht]
\includegraphics[width=\textwidth]{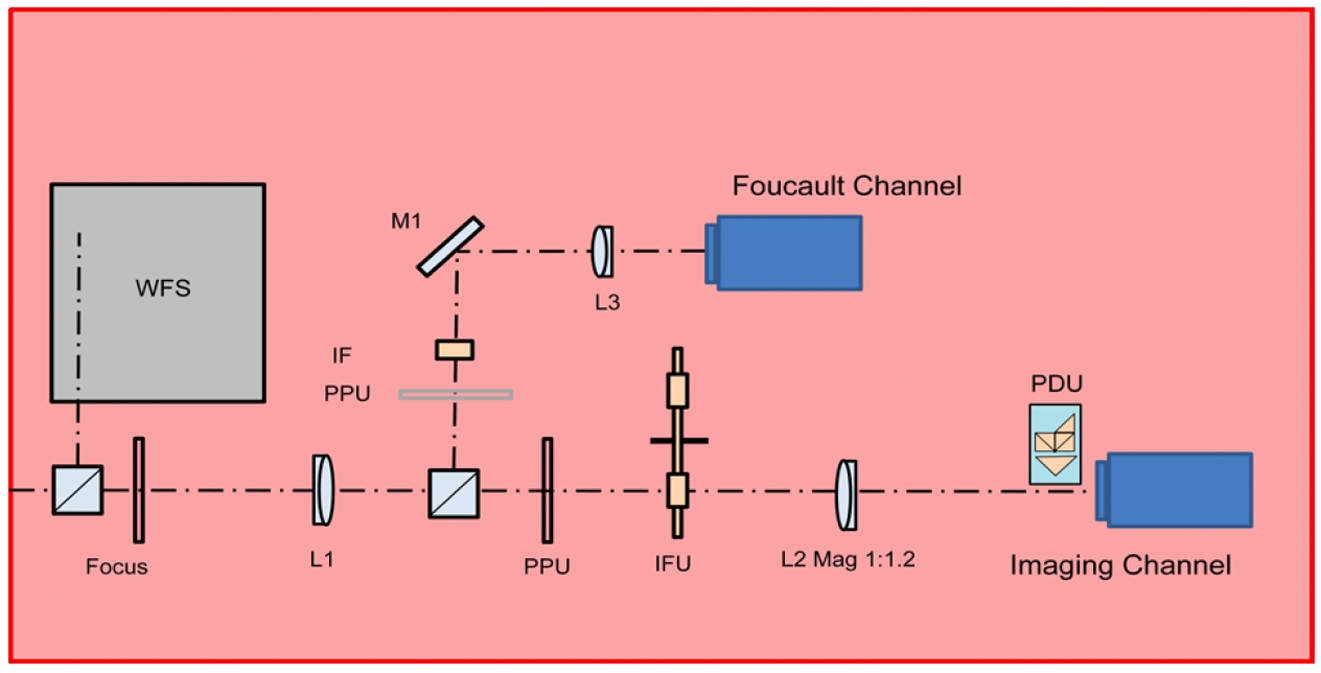}
\caption{\label{bbi_layout}Conceptual layout of the Broad-Band Imager (BBI). L1, L2: collimating and reimaging lenses, PPU: pupil plane unit, IFU: interference and neutral density filter unit, IF: interference and ND filter(s), PDU: phase diversity unit.}
\end{figure}

\subsection{Multi-wavelength measurements}
The focal-plane instruments are arranged such that at least two of them can be used simultaneously, and without compromising the photon flux to the respective instrument. The Grating Infrared Spectrograph (GRIS, see Section \ref{GRIS} below) and the Fabry-Perot-Interferometer (GFPI, see Sect. \ref{GFPI} below) are fed via a dichroic beam splitter (Fig.\ref{fig_bs}) that transmits wavelengths longer than 850 nm to the spectrograph, and reflects all shorter wavelengths to the GFPI. This allows using these existing instruments simultaneously, and in addition, it allows to add a \textit{blue} channel to the GFPI in the future, again without light loss for the other instruments involved. Similar setups are possible for the Broad-Band Imager (BBI, see Sect. \ref{bbi}) and any future instrument that would be located nearby. The arrangement of instruments is depicted in Fig. \ref{gregorlab} that shows the floor plan of the observing room. The light beam coming from the telescope is fed to the adaptive optics that is mounted on a vertical optical bench. After passing the AO system, a  rotating folding mirror \textsf{FM1} deflects the beam either to the GRIS/GFPI instrument suite, or the the BBI. The GRIS instrument is actually located one floor below the observing room, only the slit unit and the slit-jaw imaging system are located in the observing room. Light is directed from the spectrograph slit area to the main spectrograph via the folding mirror FM2.

\subsection{Broad-band imager}
\label{bbi}
\begin{figure}[!ht]
\includegraphics[width=\textwidth]{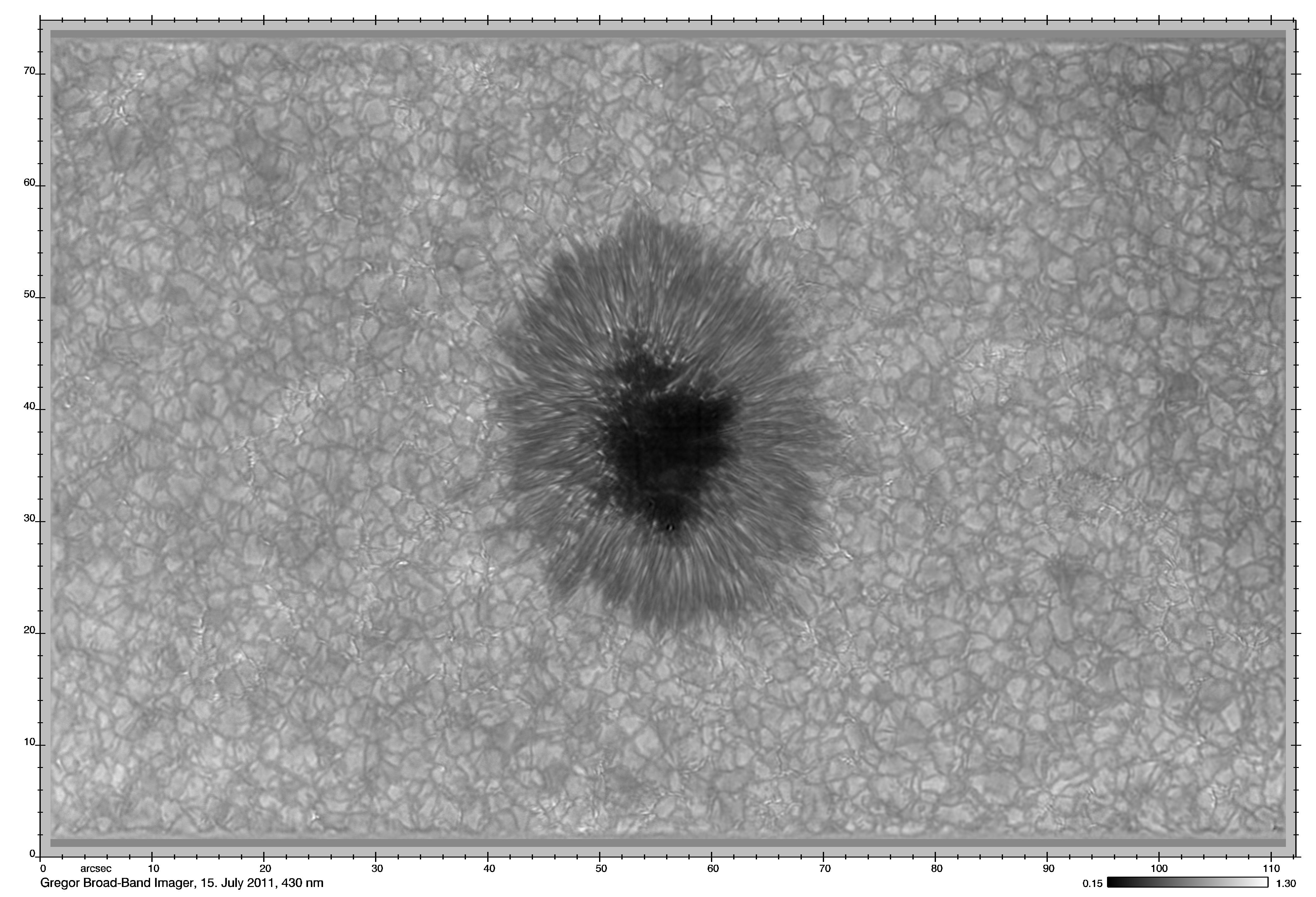}
\caption{\label{fig_g-band}Sunspot and surrounding granulation recorded on 15 July 2011 with the Broad-Band Imager of GREGOR. The data were taken at 430 nm, and a speckle reconstruction algorithm was applied. Axis labels are given in seconds of arc.}
\end{figure}

A Broad-Band Imager (BBI) has been setup at GREGOR to record high-quality images sequentially at several wavelengths, with a large field-of-view and fast cadence to allow for post-facto speckle reconstruction. Figure \ref{bbi_layout} shows the layout of the instrument. Light from the telescope is first split between the wavefront sensor of the AO and the BBI. Collimating and reimaging lenses, L1, and L2, reproduce the focal plane at a scale that allows for diffraction-limited imaging at all visible wavelengths. L1 produces a small pupil image, close to which a beam splitter reflects part of the light to the optional Foucault channel. At the location of the pupil, an aperture stop with the size of the pupil is placed.  Images are recorded with a PCO-4000 camera that has a chip size of 24$\times$36 mm$^2$ and 4008$\times$2672 square pixels with a size of 9 $\mu$.  The plate scale is 0.025 arcsec/pixel, resulting in a field of view of 102$\times$68 arcsec$^2$ (see Fig. \ref{fig_g-band}). An optional Phase Diversity (PD) unit can be placed in front of the detector. This allows PD reconstruction to be included in the toolbox for image post-processing and image quality verification. The BBI is also equipped with a Foucault channel to  measure the optical quality of the telescope. 

\subsection{GREGOR Fabry-P\'erot Interferometer (GFPI)}
\label{GFPI}
The GFPI has evolved from the G\"ottingen FPI \citep{Bendlin+etal1992, Bendlin+Volkmer1995} that had been in use at the German VTT since the early 90s. The upgrade to larger etalons, new cameras and control software, and the new optical layout for installing the GFPI at GREGOR has been described in \citet{Puschmann2006}. \citet{Bello+Kneer2008} present the full-Stokes polarimeter before the GFPI was moved to GREGOR \citep{Denker+etal2010}.
 
\begin{figure}[!ht]
\includegraphics[width=\textwidth]{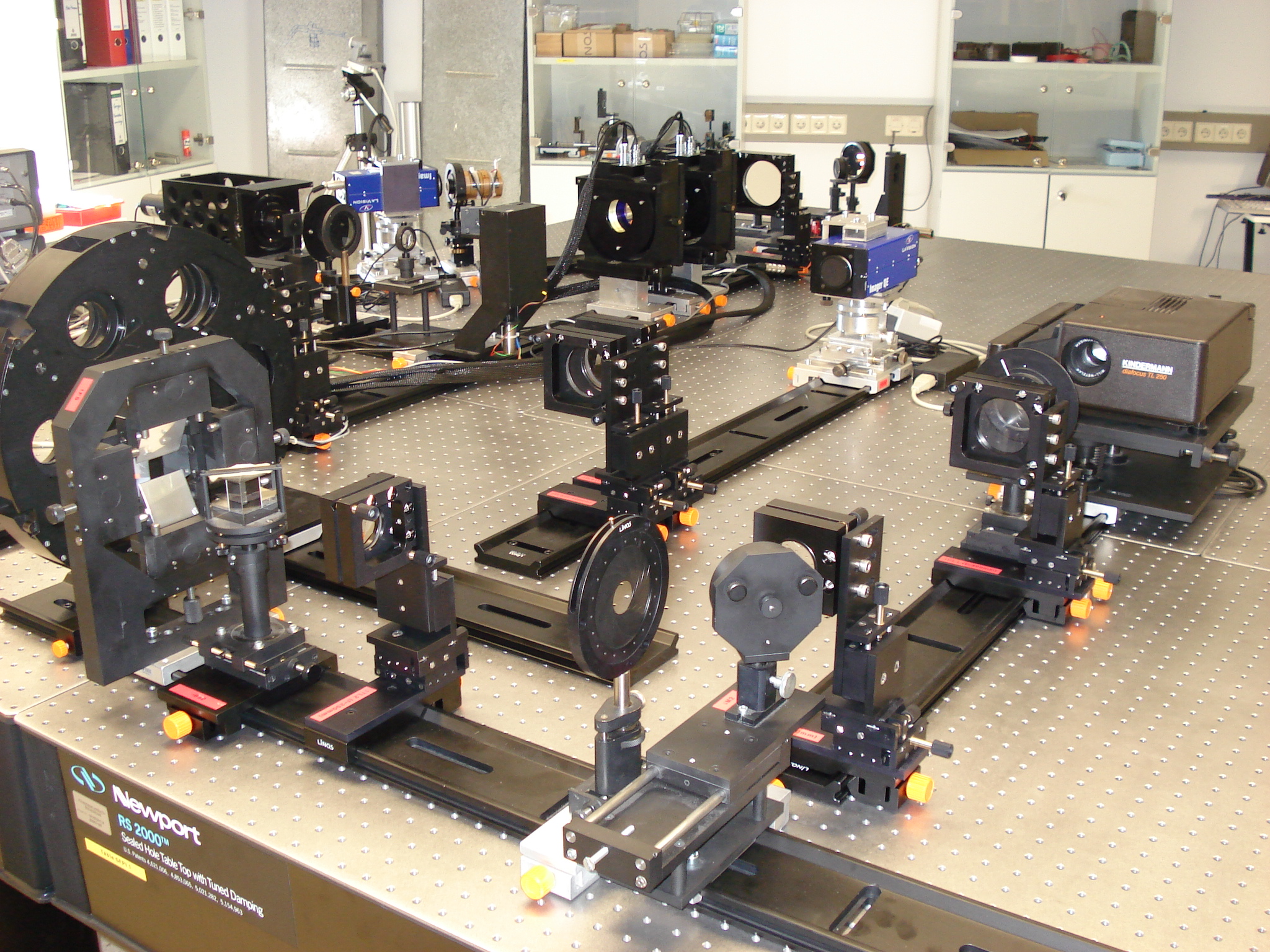}
\caption{\label{gfpi_optics}Partial view of the GREGOR Fabry Perot Instrument in the observing room. The two Etalons are located in the upper center of the picture.}
\end{figure}

The main optical paths include the narrow- and broad-band channel, as depicted in Fig. \ref{gfpi_optics}. A beamsplitter transmits 95\% of the incoming light to the narrow-band channel, while the remaining 5\% are directed to the broad-band channel. Two interference filters of 10 nm FWHM mounted on a translation stage are used to select the wavelength of the broad-band channel.

The spectroscopic channel consists of two narrow-band interference filters (FWHM 0.3 nm)  housed in a precision translation mount, followed by two tunable Fabry-P\'erot etalons with free apertures of 70 mm. The etalons are optimized for wavelengths between 530 nm and 860 nm. This range includes those photospheric lines most interesting for magnetic field diagnostics. The spectral resolution $\triangle\lambda/\lambda$ is about 250.000. Two identical detectors are used for both channels. They have 1376$\times$1040 square pixels with a pitch of 6.45$\mu$ and a full-well capacity of 18,000 electrons. The field-of-view of the detectors is 52\arcsec$\times$40\arcsec. In its polarimetric dual-beam mode, the field shrinks to 24\arcsec$\times$38\arcsec, with an image scale of 0.04\arcsec/pixel.

A dual-beam full Stokes polarimeter is an integral part of the GFPI. The polarization modulation is performed with a pair of ferro-electric liquid crystal retarders. A polarizing beam splitter allows two orthogonal polarization states to be recorded simultaneously on a single detector. This approach reduces the sensitivity of the measurements to atmospheric seeing. Four modulation states are recorded for each wavelength point. They represent a linear combination of the four components of the Stokes vector. The Stokes components  (I, Q, U, V) are obtained through demodulation as a part of the data processing pipeline \citep{Balthasar2011}.

Calibration of the instrumental polarization is an important part of the data processing. Most of this calibration is performed with a Polarization Calibration Unit \citep{Hofmann2009} that is placed in the symmetric part of the optics, between the secondary and the tertiary mirror (see Sect. \ref{design}). Only the (time-independent) polarimetric properties of the first two telescope mirrors have to be determined independently.

\subsection{GREGOR Grating Infrared Spectrograph (GRIS)}
\label{GRIS}

\begin{figure}[!ht]
\includegraphics[width=\textwidth]{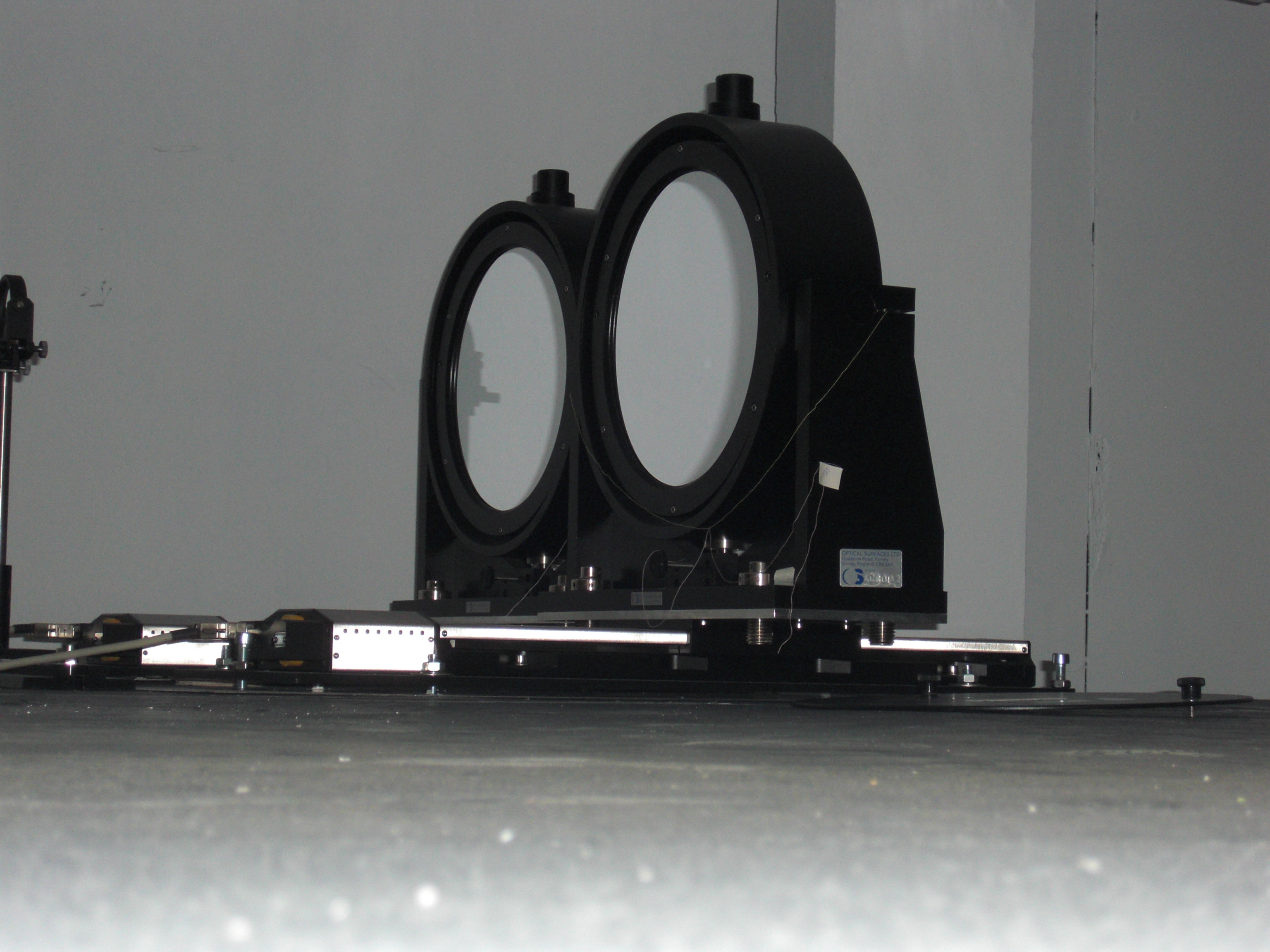}
\caption{\label{fig_gris}Collimating and imaging mirror of GRIS}
\end{figure}

The near-infrared scanning spectro-polarimeter is derived from the Tenerife Infrared Polarimeter (TIP) that was installed at the Echelle spectrograph of the VTT since 1999 \citep{Collados2008}. Im order to reduce cost, the GRIS uses existing components from TIP, namely the cryogenic IR detector and its dewar. GRIS uses the horizontal Czerny-Turner spectrograph that is located below the observing room. The corresponding collimating and imaging mirrors are shown in Fig. \ref{fig_gris}. The instrument control software and the data calibration software are mostly identical to the TIP versions, which will simplify the use of the new instrument by a great deal. GRIS has been installed during 2011, and is ready to be used in intensity mode in early 2012. Spectropolarimetric observations will be possible in the course of 2012, after the polarization model for the telescope and the instrument has been established. 

\section{Summary and Outlook}
GREGOR, Europe's largest solar telescope, will play a major role for the future of solar physics, in Europe and in the world. In Europe, it will be a unique place for high-resolution precision spectro-polarimetry at visible and infrared wavelengths. With its multi-wavelength capabilities it is an excellent facility to study the coupling of the solar atmosphere from the deepest photosphere to the upper chromosphere. GREGOR will also serve as a pathfinder for the next-generation European Solar Telescope. 

\acknowledgements 
The 1.5-meter GREGOR solar telescope is built and operated by the German consortium of the Kiepenheuer-Institut f\"ur
Sonnenphysik in Freiburg, the Leibniz-Institut f\"ur Astrophysik Potsdam, and the Max-Planck-Gesellschaft in M\"unchen with
contributions by the Institut fŸr Astrophysik der Georg-August Universit\"at G\"ottingen and other German and international partners at the Spanish Observatorio del Teide on Tenerife. The GREGOR project benefits from the skills and the dedication of the involved engineers and technicians, namely Andreas Fischer, Clemens Halbgewachs, F. Heidecke (KIS), H. Nicklas (IAG), and E. Popow (AIP).

%\bibliography{GREGOR_ws}

\end{document}